\documentclass[amsmath,amssymb,aps,citeautoscript,superscriptaddress,twocolumn,showpacs,10pt]{revtex4-2}

\usepackage{bm}

\usepackage{caption}
\usepackage{subcaption}
\usepackage{graphicx}% Include figure files
\usepackage{dcolumn}
\usepackage[colorlinks,linkcolor=red,anchorcolor=green,
citecolor=blue,breaklinks]{hyperref}

\usepackage{array, booktabs}
\usepackage{multirow}

\usepackage{mathrsfs}
\usepackage{color}
\usepackage{braket}
\usepackage[normalem]{ulem}

\graphicspath{{figures/}}
\begin{document}
\preprint{}

\title{Accurate quantum simulation of molecular ground and excited states with a transcorrelated Hamiltonian}

\author{Ashutosh Kumar}
\email{akumar1@lanl.gov}
\affiliation{Theoretical Division, Los Alamos National Laboratory, Los Alamos, NM 87545, USA}

\author{Ayush Asthana}
\affiliation{Department of Chemistry, Virginia Tech, Blacksburg, VA 24061, USA}

\author{Conner Masteran}
\affiliation{Department of Chemistry, Virginia Tech, Blacksburg, VA 24061, USA}

\author{Edward F. Valeev}
\affiliation{Department of Chemistry, Virginia Tech, Blacksburg, VA 24061, USA}

\author{Yu Zhang}
\email{zhy@lanl.gov}
\affiliation{Theoretical Division, Los Alamos National Laboratory, Los Alamos, NM 87545, USA}

\author{Lukasz Cincio}
\affiliation{Theoretical Division, Los Alamos National Laboratory, Los Alamos, NM 87545, USA}

\author{Sergei Tretiak}
\affiliation{Theoretical Division, Los Alamos National Laboratory, Los Alamos, NM 87545, USA}
\affiliation{Center for Integrated Nanotechnologies, Los Alamos National Laboratory, Los Alamos, NM 87545, USA}

\author{Pavel A. Dub}
\email{pdub@lanl.gov}
\affiliation{Chemistry Division, Los Alamos National Laboratory, Los Alamos, NM 87545, USA}

%\date{\today}
\begin{abstract}
NISQ era devices suffer from a number of challenges like limited qubit connectivity, short coherence times and sizable gate error rates. Thus, quantum algorithms are desired that require shallow circuit depths and low qubit counts to take advantage of these devices. We attempt to reduce quantum resource requirements for molecular simulations on a quantum computer, a promising application on NISQ devices, while maintaining the desired accuracy with the help of classical quantum chemical theories of canonical transformation and explicit correlation. In this work, compact \textit{ab initio} Hamiltonians are generated classically through an approximate similarity transformation of the Hamiltonian with a) an explicitly correlated two-body unitary operator with generalized pair excitations that remove the Coulombic electron-electron singularities from the Hamiltonian and b) a unitary one-body operator to efficiently capture the orbital relaxation effects required for accurate description of the excited states. The resulting transcorelated Hamiltonians are able to describe both ground and excited states of molecular systems in a balanced manner. Using the fermionic-ADAPT-VQE method based on the unitary coupled cluster with singles and doubles (UCCSD) ansatz and only a minimal basis set (ANO-RCC-MB), we demonstrate that the transcorrelated Hamiltonians can produce ground state energies comparable to the much larger cc-pVTZ basis. This leads to a potential reduction in the number of required CNOT gates by more than three orders of magnitude for the chemical species studied in this work. Furthermore, using the qEOM formalism in conjunction with the transcorrelated Hamiltonian, we reduce the errors in excitation energies by an order of magnitude. The transcorrelated Hamiltonians developed here are Hermitian and contain only one- and two-body interaction terms and thus can be easily combined with any quantum algorithm for accurate electronic structure simulations.
\end{abstract}

\maketitle

\section{Introduction}
In 1982, Feynman envisioned the idea of simulating quantum mechanical processes occurring in nature through devices that operate on the principles of quantum mechanics themselves~\cite{Feynman1982}. Since then, a lot of important progress has been made in the development of such quantum mechanical devices, also referred to as quantum computers as they promise a near-exponential speed-up (``quantum advantage'') over classical computers for a wide variety of computational tasks\cite{nielsen2002quantum}. Solving the many-body electronic Schr\"{o}dinger equation is quite naturally one of the most promising applications for quantum computers. A number of algorithms based on quantum phase estimation (QPE)\cite{abrams1999quantum, aspuru2005simulated}, adiabatic state preparation\cite{farhi2001quantum, babbush2014adiabatic}, variational quantum optimization\cite{peruzzo2014variational, o2016scalable} have been developed and refined to calculate the ground and low-lying excited states of the many-body systems with the aim of realizing the promised ``quantum advantage'' in the field of electronic structure theory. The contemporary quantum hardware, however, is still in its infancy and suffers from a variety of challenges, such as limited qubit connectivity, short coherence times and sizable gate error rates. Furthermore, the mostly one-to-one correspondence between spin-orbitals and qubits ensures that the quantum simulations can only utilize a minimal number of qubits which can, at most, only give a qualitative description of the desired solution. %\msout{However, one can't just wait for the quantum hardware to mature first, and t}
Thus, a lot of efforts lately have focused on reducing the quantum resource requirements for electronic structure simulations on noisy intermediate-scale quantum (NISQ) devices\cite{grimsley2019adaptive, tang2021qubit, fedorov2021unitary, yordanov2021molecular, adapt_vqe_rdm_gs_es_2021, tkachenko2021correlation, kottmann2021reducing, motta2021low, Bauman_downfolding_UCC_2019, Bauman_downfolding_UCC_ES_2019, clusterVQE}. A majority of these efforts utilize the variational quantum eigensolver (VQE) algorithm in conjunction with unitary coupled-cluster based ansatzes\cite{o2016scalable, peruzzo2014variational,romero2018strategies,BARTLETT1989133} and are able to produce highly compact quantum circuits through a variational minimization of the expectation value of the Hamiltonian with respect to the circuit parameters, and hence by construction, are more suited for contemporary quantum hardware. Some specific examples include development of adaptive ansatzes for simulation of ground\cite{grimsley2019adaptive, tang2021qubit, fedorov2021unitary} and excited-states\cite{yordanov2021molecular, adapt_vqe_rdm_gs_es_2021}, correlation informed permutation of qubits (PermVQE)\cite{tkachenko2021correlation} or qubits clustering (ClusterVQE)\cite{clusterVQE} approaches,  construction of highly compact molecular Hamiltonians through a basis-set free formalism\cite{kottmann2021reducing} utilizing pair-natural orbital (PNO) based compression \cite{VRG:riplinger:2016:JCP, VRG:kumar:2020:JCP} in conjunction with multi-resolution\cite{MRA_Harrison_2004} strategies, low rank factorization techniques for approximating operators\cite{motta2021low}.
In a similar work, Bauman et al. employed %\msout{quantum phase estimation (}
the QPE algorithm and double unitary coupled-cluster (DUCC) formalism to downfold or embed many-body correlation effects into active-spaces of effective Hamiltonians for both ground\cite{Bauman_downfolding_UCC_2019} and excited states\cite{Bauman_downfolding_UCC_ES_2019}. The explicitly correlated theories\cite{Kutzelnigg1985R12, VRG:kong:2012:CR,Hattig:2012dz,TenNo:2012bb}, which are routinely used in classical electronic structure calculations to accelerate the convergence of electronic energies and other molecular properties with respect to the size of basis sets, is another attractive approach, which has the potential to significantly reduce the computational resources required for accurate quantum simulations. It is well known that the traditional many-body wavefunctions generated from a superposition of single slater determinants which are nothing but an antisymmetrized product of one-electron orbitals fail to capture short-range dynamic correlation effects efficiently.\cite{VRG:kong:2012:CR}. The singularity of the Coulombic electron-electron interactions near the coalescent point introduces cusps\cite{kato1957eigenfunctions} in the wavefunction, which can only be described accurately by using a large number of one-electron basis functions. The explicitly correlated methods alleviate this problem through an explicit parametrization of the wavefunction in terms of inter-electronic distances and are hence referred to as ``R12'' or ``F12'' methods. Other F12-based formalisms have also been developed which focus on removing the singularities in the Hamiltonian itself\cite{BoysHandy1969transcorrelated, Tenno2000transcorrelated, Luo2010transcorrelated}. In the transcorrelated Hamiltonian approach originally introduced by Boys and Handy\cite{BoysHandy1969transcorrelated} with later improvements by Ten-no\cite{Tenno2000transcorrelated} and Luo\cite{Luo2010transcorrelated}, singularity-free Hamiltonians are generated through a similarity transformation of the Hamiltonian with a geminal correlation operator $\hat{A}$,
\begin{equation}
\hat{H} \to \hat{\bar{H}} = e^{-\hat{A}} \hat{H} e^{\hat{A}} \quad .
\label{eq:ST}
\end{equation}
Yanai and Shiozaki\cite{yanai2012canonical} utilized ideas from both the canonical transformation theory\cite{Chan2006CanonicalTransform,Chan2010CanonicalTransform} and the transcorrelated approach to construct canonical transcorrelated (CT-F12) Hamiltonians, where unlike the previous works on the transcorrelation theory, they used a unitary geminal operator which ensures that the transformed Hamiltonian is Hermitian, making the formalism quite robust and easy to use.
% \begin{equation}
% \begin{split}
% \hat{\bar{H}} &= e^{ - \hat{A} } \hat{H} e^{ \hat{A} } \\
% &\approx \hat{H} + {[ \hat{H}, \hat{A}]}_{1,2} + \frac{1}{2} {{[[ \hat{H}, \hat{A}]}_{1,2}, \hat{A}]}_{1,2} + \dots \; .
% \label{eq:CT}
% \end{split}
% \end{equation}
%use of a unitary geminal operator, truncation of the Baker–Campbell–Hausdorf expansion etc.\cite{} to construct canonical transcorrelated F12 (CT-F12) Hamiltonian which is Hermitian and contains only one- and two-body interaction terms, just like the untransformed Hamiltonian. 
In an earlier work by some of us\cite{Motta_CTF12_quantum}, the CT-F12 Hamiltonian, in conjunction with the UCCSD ansatz and VQE algorithm was able to produce near cc-pVTZ quality ground state correlation energies of several small molecular species, with the much smaller 6-31g basis. McArdle and Tew\cite{mcardle2020improving} and more recently, Sokolov and co-workers\cite{sokolov2022orders}, also employed the transcorrelated approach to improve the accuracy of quantum simulations, where they had to make use of the imaginary-time evolution algorithms due to the non-Hermiticity of their transformed Hamiltonian. In another application of explicit correlation strategies in quantum computing, Schleich and co-workers\cite{schleich2021improving} 
recently employed the $[2]_{R12}$ formalism developed by Valeev and Torheyden\cite{torheyden2009universal},
where one- and two-body reduced density matrices obtained from the quantum simulation was used to formulate
a correction to the energy. However, these \textit{a posteriori} corrections have only been developed for ground-state correlation energies, while \textit{a priori} strategies like CT-F12 can be potentially combined with any many-body quantum theory to calculate different molecular properties corresponding to both ground and excited states.\\ In this work, we look to extend the CT-F12 formalism to the simulation of molecular excited states as many interesting chemical phenomena in nature involve excited states in one way or the other.
Preliminary investigations with the CT-F12 Hamiltonian revealed a very strong bias towards the ground-state and the basis-set convergence of excited-state properties like excitation energies was noticeably slower than the regular Hamiltonian itself. Similar observations were also noticed in the framework of explicitly correlated coupled-cluster response theory \cite{CCR12_ES}. One reason for this unbalanced description of the ground and excited states can be attributed to the absence of virtual orbitals in the definition of the geminal operator. This is quite obvious in the case of valence excited states where an accurate simulation would require the inclusion of dynamic correlation effects between an electron in occupied and virtual orbital in the effective Hamiltonian. For example, the ${}^1P (2p \leftarrow 2s)$ state of the Be atom would be very poorly described with the above formalism as both the occupied orbitals (1s and 2s) are of S symmetry and hence the resulting pair won't contribute at all to the given excited state\cite{CC_R12_Be}. Furthermore, the basis-set convergence of the energies of Rydberg-like excited states are dominated by non-dynamical electron correlation effects, and one needs to incorporate orbital relaxations in the effective Hamiltonian for their accurate description. 

In this work, we have added pairs involving virtual orbitals in the definition of the geminal operator along with the introduction of a singles operator in the similarity transformation procedure. It should be noted that Watson et al.\cite{WatsonChanMB} had implemented a similar approach for accurate molecular computations on a classical computer with a minimal basis. In this paper, we develop an improved recipe for obtaining the singles operator (see sec.~\ref{sec:theory} for details) and apply the method to achieve a massive reduction in quantum resources for the molecular ground and excited state calculations on a quantum computer.
Within the framework of VQE algorithms, different strategies have been developed to compute the energies of the low-lying excited states of molecular species\cite{peruzzo2014variational, Lee_OC_VQE_2019}. State-specific methods, that compute one excited state at a time, include folded spectrum method \cite{peruzzo2014variational} and orthogonality constrained VQE method\cite{Lee_OC_VQE_2019}. A more general approach, Quantum Krylov subspace expansion methods\cite{motta2020determining, stair2020multireference, cortes2021quantum,yeter2021benchmarking} diagonalize the Hamiltonian in a small subspace and can provide a number of low-lying excited states together. In addition, subspace expansion based on excited determinants has been proposed\cite{colless2018computation,parrish2019quantum}, including equation of motion (EOM) operator based qEOM method\cite{pauline_qeom_2020} that provides a size-intensive approach for excitation energy calculations. In order to demonstrate the advantage of transcorrelated methods in molecular simulations on a quantum computer, we use a classical simulator to compute the ground state energies through fermionic-ADAPT-VQE\cite{grimsley2019adaptive} and excitation energies using the formalism of qEOM implemented on top of the ground state calculation.  \\
This manuscript is organized as follows: Section \ref{sec:theory} introduces the formalism of the canonical transcorrelation procedure. Computational details are provided in Sec.~\ref{sec:comp_details}. In Sec.~\ref{sec:results}, we assess the 
accuracy of the transcorrelated Hamiltonian in calculations of ground and excited states energies of a number of small molecular species and estimate the potential reduction in quantum resources with this approach. We give a summary of our findings in Sec.~\ref{sec:summary}.

\section{Theory}\label{sec:theory}
\begin{figure}
    \includegraphics[width=0.95\columnwidth]{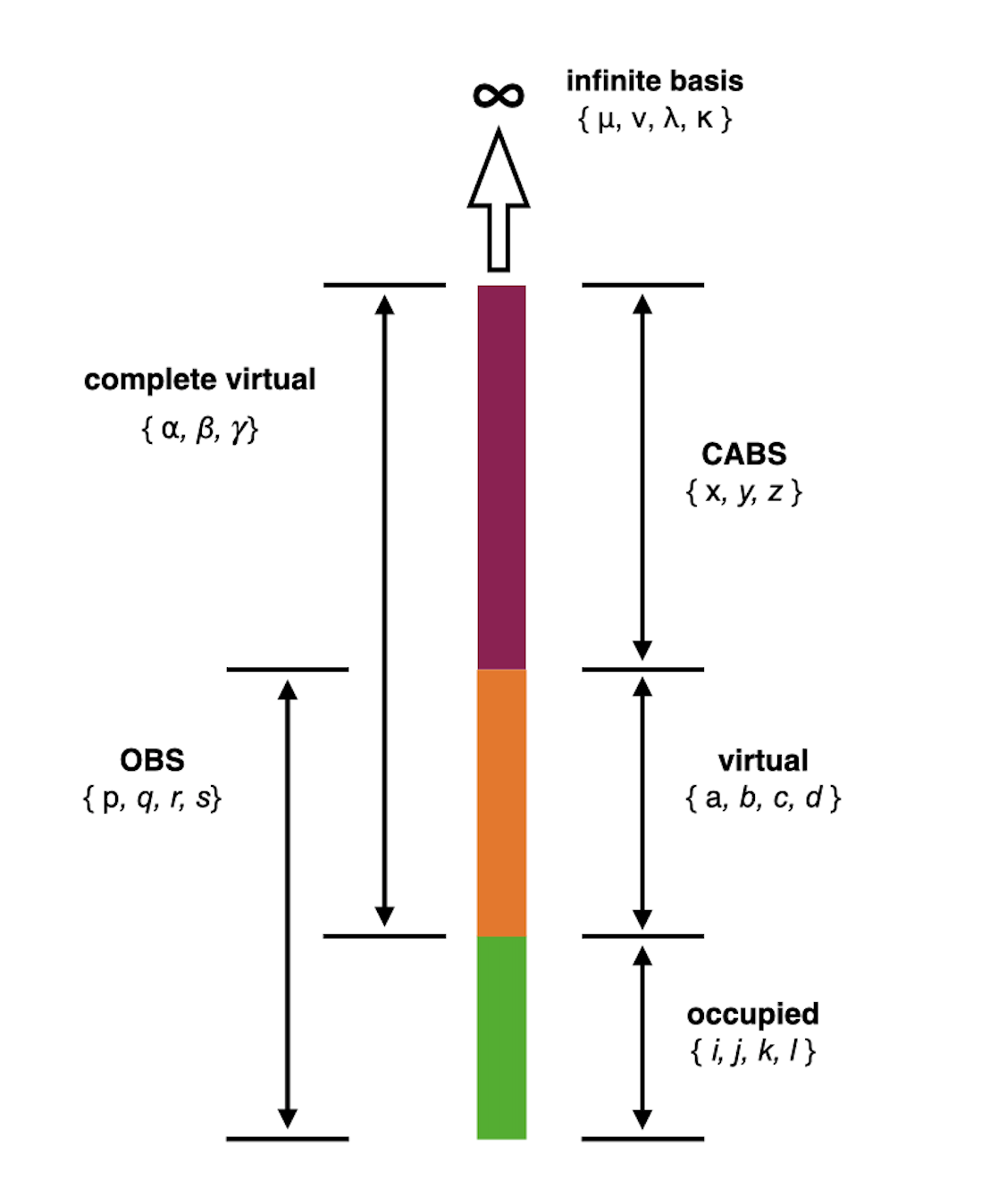}
    \caption{Schematic notation used for different orbital spaces. OBS here refers to 
    the finite orbital basis while CABS is the complementary auxiliary basis set, the 
    orthogonal complement to the OBS space.}
    \label{fig_spaces}
\end{figure}

The molecular Hamiltonian in the second-quantized formalism can be written as,
\begin{equation}
\hat{H}=h_{\nu}^{\mu} \hat{E}_{\mu}^{\nu}+\frac{1}{2} g_{\nu \kappa}^{\mu \lambda} \hat{E}_{\mu \lambda}^{\nu \kappa}
\label{eq:H}
\end{equation}
where $h_{\nu}^{\mu}$ and $g_{\nu \kappa}^{\mu \lambda}$ refer to the one and two-electron elements of the Hamiltonian
and {$\mu$, $\nu$, $\kappa$, $\lambda$} indices refer to the orbitals in the infinite orbital basis. Please refer 
to Fig.~\ref{fig_spaces} for a detailed description of the orbital spaces along with their labels, used in this 
work. Here, $\hat{E}_{\mu}^{\nu}$ is the spin-free or spin-summed excitation operator, %$E_{\mu}^{\nu}=\sum_{\sigma \in\{\alpha \beta\}} a_{\mu \sigma}^{\dagger} a_{\nu \sigma}$, 
$E_{\mu}^{\nu}=a_{\mu \sigma}^{\dagger} a_{\nu \sigma}$, 
where $a_{\mu \sigma}^{\dagger}$
and $a_{\mu \sigma}$ ($\sigma \in\{\alpha \beta\}$) are the usual creation and annihilation operators respectively, with $\sigma$ referring to the the spin label. We have followed the Einstein summation convention throughout this work.
The canonical transcorrelated (CT) theory\cite{yanai2012canonical} aims to incorporate the missing dynamic electron correlation effects into an effective Hamiltonian through a similarity transformation, $\hat{\bar{H}}=e^{\hat{A}^{\dagger}} \hat{H} e^{\hat{A}}$, where $\hat{A}$ is usually an anti-Hermitian many-body operator which makes $e^{\hat{A}^{\dagger}}$ and $e^{\hat{A}}$ operations unitary, thus maintaining the Hermiticity of the Hamiltonian. Utilizing the Baker–Campbell–Hausdorff (BCH) expansion, the effective Hamiltonian can be expressed in terms of (nested) commutators,
\begin{equation}
\hat{\bar{H}}=\hat{H}+[\hat{H}, \hat{A}]+\frac{1}{2 !}[[\hat{H}, \hat{A}], \hat{A}]+\ldots .   
\end{equation}
The CT theory introduces additional approximations of a) restricting the above expansion to only double commutators and b) approximating the full Hamiltonian ($\hat{H}$) by its mean-field constituent, the Fock operator($\hat{F}$) in the double commutator term,
 \begin{equation}
\hat{\bar{H}}
\approx
\hat{H}
+ {[ \hat{H}, \hat{A}]}_{1,2}
+ \frac{1}{2} {{[[ \hat{F}, \hat{A}]}_{1,2}, \hat{A}]}_{1,2}
\quad.
\label{eq:CT-F12}
\end{equation}
The resulting CT-F12 Hamiltonian is correct at least through second-order in perturbation\cite{kutzelnigg1992stationary}. It should be noted that similar approximations have also been used in CC-F12 theories as well\cite{CCSD_F12_2010}. The notation $[..]_{1,2}$ means only one- and two-body operators generated from the given commutator are retained directly while the three-body operators are included through an approximate decomposition into one- and two-body operators using the extended normal ordering approach of Mukherjee and Kutzelnigg\cite{mukherjee1997normal,kutzelnigg1997normal,Kutzelnigg_MK_ENO}. In spin orbitals representation, the full decomposition can be written as\cite{TakeshiChan_CT},
\begin{equation}
\begin{aligned} 
a_{s t u}^{p q r} \hspace{0.01 in} = \hspace{0.05 in} & \tilde{a}_{s t u}^{p q r} + 9\left(D_{s}^{p} \wedge a_{t u}^{q r}\right)-36\left(D_{s}^{p} \wedge D_{t}^{q} \wedge D_{u}^{r}\right) \\& +9\left(D_{s t}^{p q} \wedge a_{u}^{r}\right) +24\left(D_{s}^{p} \wedge D_{t}^{q} D_{u}^{r}\right)-9\left(D_{s t}^{p q} \wedge D_{u}^{r}\right)\\
&+ \lambda^{pqr}_{stu},
\end{aligned}
\end{equation}
% eq. 25 of [Canonical transformation theory from extended normal ordering]
where $\wedge$ denotes antisymmetrization over all upper and lower indices\cite{TakeshiChan_CT} with the corresponding prefactor of $(\frac{1}{n})^2$, where $n$ is the particle rank of the original undecomposed operator (n = 6 for three-body decompositions). In this work, the first term $\tilde{a}_{s t u}^{p q r}$, which is the three-body fluctuation operator in normal ordered form with respect to a reference and the last term $\lambda^{pqr}_{stu}$ which refers to the three-body density cumulants, have been dropped from the above decomposition. The final spin-free equations for this approximate decomposition can be found in Ref.\cite{Kutzelnigg_MK_ENO}. The mean-field one-body Fock operator in equation 3. 
is defined as, %\yz{change $v$ to $\nu$, make it consistent}
\begin{equation}
\begin{split}
    & \hat{F}=f_{\nu}^{\mu} \hat{E}_{\mu}^{\nu},\\
    & f_{\nu}^{\mu}=h_{\nu}^{\mu}+D_{\kappa}^{\lambda}\left(g_{\nu \kappa}^{\mu \lambda}-\frac{1}{2} g_{\kappa \nu}^{\mu \lambda}\right)
\end{split}    
\end{equation}
where, $D_{\nu}^{\mu}=\left\langle\Psi_{0}\left|\hat{E}_{\nu}^{\mu}\right| \Psi_{0}\right\rangle$ is the one-body reduced density matrix associated with the reference wavefunction $\Psi_{0}$.
We have used the following form of the transformation operator $\hat{A}$ in this work,
\begin{equation}
\begin{split}
& \hat{A} = \hat{A}^{\mathrm{F} 12 } + \hat{S'},\\
& \hat{A}^{\mathrm{F}12} =  \frac{1}{2} G_{p q}^{\alpha \beta}\left(\hat{E}_{p q}^{\alpha \beta}-\hat{E}_{\alpha \beta}^{p q}\right),\\
& \hat{S'} = G_{p}^{\alpha}\left(\hat{E}_{p}^{\alpha} - \hat{E}_{\alpha}^{p}\right).
%\hat{A}^{\mathrm{F} 12 }= G_{p}^{\alpha}\left(\hat{E}_{p}^{\alpha} - \hat{E}_{\alpha}^{p}\right) 
%+ \frac{1}{2} G_{p q}^{\alpha \beta}\left(\hat{E}_{p q}^{\alpha \beta}-\hat{E}_{\alpha \beta}^{p q}\right),
\end{split}
\end{equation}
The amplitudes corresponding to the geminal operator $\hat{A}^{\mathrm{F}12}$ is defined as,
\begin{equation}
G_{p q}^{\alpha \beta} = \frac{3}{8}\left\langle\alpha \beta\left|\hat{Q}_{12} \hat{F}_{12}\right| p q \right\rangle
+\frac{1}{8}\left\langle\alpha \beta\left|\hat{Q}_{12} \hat{F}_{12}\right| q p\right\rangle.
\end{equation}
Here, we have made use of the SP ansatz\cite{TenNo2004SPAnsatz,Zhang:2012it} of Ten-no where the geminal amplitudes (1/8, 3/8) are fixed and are obtained by satisfying the first-order cusp conditions for the singlet and triplet electron pairs respectively. We also chose a slater-type geminal (STG) as the two-body
%(\yz{two-body was used before, make the terminology consistent throughout the paper}) 
correlation factor,
\begin{equation}
\hat{F}_{12}(r_{12})=-\gamma^{-1} \exp \left(-\gamma r_{12}\right),
\end{equation}
where $\gamma$ is a scale-length parameter whose values are in practice tuned to a given orbital basis set\cite{Klopper2005Gamma}. 
The strong orthogonality projector $\hat{Q}_{12}$,
\begin{equation}
\hat{Q}_{12}=1-\hat{V}_{1} \hat{V}_{2},
\end{equation}
where $\hat{V}_i$ projects the one-electron states into virtual orbitals (a,b) of the orbital basis set (OBS), ensures that geminal matrix elements involving products of virtual orbitals like $\langle ab | \hat{Q}_{12} \hat{F}_{12}| pq\rangle = 0$. Thus, all the geminal matrix elements considered here contain at least one external (CABS) index. \\
In an earlier work by some of us\cite{Motta_CTF12_quantum}, only occupied-occupied pairs were included in the definition of the geminal operator. This made the transcorrelated Hamiltonian very biased towards the ground state wavefunction. The orbital pairs involving the virtual orbitals, especially the chemically important ones (usually defined in active spaces) can contribute very significantly to the excited state wavefunction and thus one needs to satisfy the cusp conditions for such pairs as well. Thus, we included the occupied-virtual and virtual-virtual geminal pairs as well in this work, so that our Hamiltonian can treat multiple states at an equal footing. This approach is very common in quasi-degenerate perturbation theory\cite{nakano2002quasi} where a multi-configuration reference wavefunction is deployed instead of the regular Hartree-Fock wavefunction to remove any kind of biasedness from the Hamiltonian. However, not all virtual orbitals are equally
important and one can always generate an active space instead. One way to do this is to look at the eigenvalues (also called occupation number) of the virtual-virtual block of the MP2 one-body reduced density matrix\cite{CCR12_ES}.
%(\yz{is it similar to the natural orbital analysis for selecting active space?})
and choose only those natural virtual orbitals with occupation numbers greater than a given threshold.
\\Furthermore, we have also added a singles operator in the similarity transformation procedure to incorporate orbital relaxation effects in the Hamiltonian. In the equation of motion based formalisms, the excited state wavefunction is often characterized by dominant  contributions from the singles excitation operator (for example $R^a_i$ amplitudes in EOM-CCSD) and thus addition of quality singles amplitudes is essential. In order to define our singles operator, we look towards the ``CABS singles'' approach usually employed in the explicit correlation theory to accelerate the basis set convergence of the energy of the reference wavefunction by allowing for orbital rotations between the occupied and the missing virtual space i.e. the CABS space. Following the works of Valeev and Kong\cite{VRG:kong:2012:CR}, these amplitudes have been determined from the Rayleigh-Schr\"{o}dinger perturbation theory, with the following 
partitioning of the Hamiltonian,
\begin{equation}
    \hat{H}^{(0)}=\left(\begin{array}{ccc}F_{j}^{i} & 0 & 0 \\ 0 & F_{b}^{a} & F_{b}^{x} \\ 0 & F_{y}^{a} & F_{y}^{x}\end{array}\right), \hspace{0.05 in} \hat{H}^{(1)}=\left(\begin{array}{ccc}0 & F_{i}^{a} & F_{i}^{x} \\ F_{a}^{j} & 0 & 0 \\ F_{y}^{j} & 0 & 0\end{array}\right)
\end{equation}
\\where $F^{i}_{j}$, $F^{a}_{b}$, $F^{x}_{i}$ and $F^{y}_{x}$ refer to the occupied-occupied, virtual-virtual, occupied-CABS and CABS-CABS block of the Fock matrix respectively. In this approach, the occupied-CABS ($G^x_i$) block of the singles amplitudes can be obtained by solving the following equation,
\begin{equation}
    F^{j}_{i} G^{x}_{j} - F^{x}_{y} G^{y}_{i} = F^{x}_{i}.
\end{equation}
We chose the perturbative formulation of the ``CABS singles'' approach in this work as it has been shown to work quite well for small basis sets\cite{VRG:kong:2012:CR}. One can easily extend the above equation to solve for the virtual-CABS ($G^x_a$) component of the singles amplitudes as well but this naive approach often leads to very poor results as the resulting equations are generally very poorly conditioned due to the small differences in orbital energies ($F^{\mu}_{\mu}$) between virtual and CABS orbitals. In order to overcome this, we replace the virtual-virtual block of the Fock matrix by a constant parameter $\epsilon$ which we denote as ``shift'' in this work,
\begin{equation}
    \epsilon * G^{x}_{a} - F^{x}_{y} G^{y}_{a} = F^{x}_{a}.
\end{equation}
This shift parameter is in practice very close to the HOMO energy but can be adjusted for a given molecule and basis set to add optimal orbital relaxation effects in the transcorrelated Hamiltonian. This concept is very similar to the regularization strategies employed in the single reference-based perturbation theories to treat multi-reference problems\cite{2018LeeROOMP2}.
In other approximations, following Shiozaki and Yanai,\cite{yanai2012canonical} $G^{cx}_{ab}$ types of geminal amplitudes have been ignored as incorporating them would require addition of full three body operators and not the approximated ones used in this work.\\ Finally, the transcorrelated Hamiltonian has the following form,
\begin{equation}
    \hat{\bar{H}}=\bar{h}_{q}^{p} \hat{E}_{p}^{q}+\frac{1}{2} \bar{g}_{r s}^{p q} \hat{E}_{p q}^{r s},
\end{equation}
where $\bar{h}_{q}^{p}$ and $\bar{g}_{r s}^{p q}$ are the ``perturbed'' one- and two-body interaction terms. The transcorrelated Hamiltonian ($\hat{\bar{H}}$) is Hermitian but its two-body interaction term has a lower symmetry compared to the original Hamiltonian: $\overline{g}^{pq}_{rs} \neq \overline{g}^{ps}_{rq}$ while $g^{pq}_{rs} = g^{ps}_{rq}$.
The construction of the transcorrelated Hamiltonian (with fully factorized equations) scale as $\mathcal{O}$($N^6$), where $N$ is the number of orbitals, with a quadratic dependence on the size of the CABS basis when approach C\cite{kedvzuch2005alternative} is used to evaluate the intermediate B.
\section{Computational details}\label{sec:comp_details}
The second quantized expressions required to construct the transcorrelated Hamiltonian were derived using a python module\cite{gen_no_wicks} that automates the evaluation of single and double commutators of the Hamiltonian with excitation and de-excitation operators of different orders using Wick's theorem. The resulting expressions were automatically converted to the \texttt{einsum} tensor contraction routines from the \texttt{numpy}\cite{harris2020array} library to generate the final transcorrelated Hamiltonian. All the integrals and intermediates (V, X and B)\cite{VRG:kong:2012:CR} were calculated using the MPQC\cite{MPQCPublicRepo} software. We used the approach C to evaluate the matrix elements of the B intermediate. \\ In this study, we consider four small hydrogen-containing molecules: H$_2$ \texttt{(r(H-H) = 0.7 {\AA})}, LiH \texttt{(r(Li-H) = 1.5957 {\AA})}, H$_2$O \texttt{(r(O-H) = 0.957 {\AA}, $\theta$ (H-O-H) = 104.5\textdegree)} and NH$_3$ \texttt{(r(N-H) = 1.09 {\AA}, $\theta$ (H-N-H) = 109.427\textdegree)}. 6-31g basis set was used as the OBS in the calculations involving the H$_2$ molecule while the ANO-RCC-MIN\cite{roos2005new} basis set was employed for all others. In the canonical transcorrelation procedure, aug-cc-pVTZ-OptRI\cite{yousaf2009optimized} basis set was used as the CABS basis for all the molecules studied in this work except the LiH molecule where cc-pVDZ-F12-OptRI\cite{Peterson_cabs_2008} basis set was used instead due to the non-availability of the former. All the transcorrelated calculations utilized the CABS+ approach\cite{valeev2004improving}. Since, optimized $\gamma$ values are only available for larger basis sets, we chose those $\gamma$ values for a given molecule and basis set which produced ground state energies in the close vicinity of the cc-pVTZ value. Similarly, those values of the shift parameter were chosen which minimized the maximum deviation of the excitation energies from the reference values. For a detailed analysis, we compare the performance of six types of Hamiltonians in this work. $\hat{H}$ refers to the regular untransformed Hamiltonian while $\hat{H}_\text{F12}^{(ij)}$ and $\hat{H}_\text{F12}^{(pq)}$ refer to the transcorrelated Hamiltonians generated by the doubles geminal operator defined  by occupied-occupied and all-all orbital pairs respectively. Consequently, addition of the singles operator in the similarity transformation procedure leads to S' + $\hat{H}_\text{F12}^{(ij)}$ and S' + $\hat{H}_\text{F12}^{(pq)}$ transcorelated Hamiltonians. Finally, S' + $\hat{H}$ is obtained from the similarity transformation of the regular Hamiltonian with the singles operator only.
The ground state energies associated with all these Hamiltonians were calculated using the fermionic-ADAPT-VQE method (implemented clasically\cite{adapt_vqe}) with Jordan-Wigner mapping in conjunction with the UCCSD ansatz. Excitation energies were calculated using the qEOM formalism\cite{pauline_qeom_2020} on top of the ground state calculations. The reference cc-pVTZ ground and excited state energies were calculated classically using CCSD and EOM-CCSD methods respectively using the PySCF software\cite{Pyscf_2020}

\section{Results and Discussions}\label{sec:results}
We test the performance of the transcorrelated Hamiltonians by doing quantum simulations of both 
ground and excited states on a number of small hydrogen containing molecules: H$_2$, LiH, H$_2$O and NH$_3$ 
using 6-31g and the minimal basis set ANO-RCC-MB. We compare the ground state energies and excitation 
energies of the few lowest-lying excited states of these molecules with the corresponding values obtained 
with a much larger and more accurate basis set cc-pVTZ. 
% Results for H2
%\begin{table*}{!htb}
\begin{table*}
\captionsetup{justification=raggedright}
\caption{Deviations obtained in the ground state energy (mEh) and the excitation energies (eV) %(\yz{why don't use the same unit, mEh? because some values will be too large?}) 
of the four lowest-lying excited states of the H$_2$ molecule using the 6-31g basis set from the reference cc-pVTZ values (second column) for the six different Hamiltonians (see text for details). Parameters used: ($\gamma$, shift) = (0.7, -0.4)}
\begin{ruledtabular}
\begin{tabular}{lccccccc}
{} & {} & {} & {} & {} & {} & {} & {} \\ % blank line
\multicolumn{1}{c}{States} &  \multicolumn{1}{c}{Reference} & \multicolumn{1}{c}{$\hat{H}$} & \multicolumn{1}{c}{$\hat{H}_\text{F12}^{(ij)}$} & \multicolumn{1}{c}{$\hat{H}_\text{F12}^{(pq)}$} & \multicolumn{1}{c}{S' + $\hat{H}_\text{F12}^{(ij)}$} & \multicolumn{1}{c}{S' + $\hat{H}_\text{F12}^{(pq)}$} & \multicolumn{1}{c}{S' + $\hat{H}$}\vspace{0.25cm}\\
\hline
{} & {} & {} & {} & {} & {} & {} & {}\\ % blank line
{} &  \multicolumn{1}{c}{cc-pVTZ}  &  &  &  \multicolumn{2}{c}{6-31G} &  &  \\
\cmidrule(l){3-8}
{} & {} & {} & {} & {} & {} & {} & {}\\ % blank line
S$_0$ & -1.17101 & 20.8 & 6.8 & 3.3 & 2.5 & -1.6 & 16.0\vspace{0.3 cm} \\
\hline
{} & {} & {} & {} & {}  & {} & {} & {}\\ % blank line
T$_1$ & 11.30 & 0.08  & 0.46 & 0.50 & 0.24 &  0.25  &  -0.13 \vspace{0.2 cm} \\
S$_1$ & 13.89 & 1.76  & 2.14 & 1.45 & 0.26 &  -0.15 & -0.11 \vspace{0.2 cm} \\
T$_2$ & 15.25 & 7.84  & 8.23 & 8.15 & -0.23 & -0.31 & -0.60 \vspace{0.2 cm} \\
S$_2$ & 17.60 & 10.75 & 11.14 & 10.43 & 1.01 & 0.17 &  0.62 \vspace{0.2 cm} \\
\end{tabular}
\end{ruledtabular}
\label{Tab:H2}
\end{table*}

\subsection{H$_2$}
Table~\ref{Tab:H2} lists the deviations obtained in the ground state energy (mEh) and the excitation energies (eV) of the four lowest-lying excited states (arranged in the order of increasing energies) of the H$_2$ molecule using 6-31g basis set for all the six types of Hamiltonians considered in this work, from the reference cc-pVTZ values (second column). S$_i$ and T$_i$ symbols in the first column refer to the $i^{th}$ singlet and triplet states respectively. The values of $\gamma$ and shift parameters used in these calculations were 0.7 and -0.4 respectively. The ground state energy with the regular Hamiltonian is around 21 mEh away from the reference value. While all the transcorrelated Hamiltonians bring this difference down, the  ones with the singles operator in the similarity transformation procedure, S' + $\hat{H}_\text{F12}^{(ij)}$ and S' + $\hat{H}_\text{F12}^{(pq)}$ perform the best with deviations of only 2.5 mEh and -1.6 mEh respectively. Thus, the S' + $\hat{H}_\text{F12}^{(pq)}$ Hamiltonian produces even better energies than the reference cc-pVTZ values and is in fact identical to the total energy obtained using the cc-pVQZ basis, which is quite close to the complete basis limit (CBS). Looking at the excitation energies, the $\hat{H}_\text{F12}^{(ij)}$ Hamiltonian gives larger deviations than the regular Hamiltonian itself. The deviations grow sharply as we move towards the higher energy excited states, with a max deviation of 11.14 eV observed for the S$_2$ excited state. This clearly shows that an unbalanced description of the ground and excited states is obtained when geminal operators are defined with only occupied-occupied orbital pairs. On adding all the orbital pairs in the geminal operator ($\hat{H}_\text{F12}^{(pq)}$), the max deviation is lowered but only slightly to 10.43 eV. Adding the singles operator in the similarity transformation on the other hand has quite a dramatic effect with a max deviation of 1.01 eV and 0.31 eV for the S' + $\hat{H}_\text{F12}^{(ij)}$ and S' + $\hat{H}_\text{F12}^{(pq)}$ Hamiltonians respectively. For the triplet excited states (T$_1$, T$_2$), the S' + $\hat{H}_\text{F12}^{(ij)}$ Hamiltonian performs slightly better over its $pq$ counterpart but the magnitude of improvement in excitation energies for these states are only 0.01 eV and 0.08 eV respectively. It should be noted that the correction due to the S' + $\hat{H}_\text{F12}^{(ij)}$ Hamiltonian can be seen to have a state-specific nature. It will always be biased towards those excited states with dominant contributions from configurations with the ``regular'' occupied pairs. On the other hand, the S' + $\hat{H}_\text{F12}^{(pq)}$ Hamiltonian is more reliable as it also takes care of the missing dynamic correlation effects due to electrons in occupied-virtual and virtual-virtual orbital pairs. Thus, it was able to reduce the max deviation in excitation energies from 1.01 eV (S' + $\hat{H}_\text{F12}^{(ij)}$) to 0.31 eV as seen in table ~\ref{Tab:H2}. From these results, it's quite clear that the S' + $\hat{H}_\text{F12}^{(pq)}$ Hamiltonian should be preferred for accurate simulation of both ground and excited states.\\
Furthermore, one can easily see that the basis set convergence of the energies of these excited states are not dominated by dynamical electron correlation effects and hence can't be captured by the geminal operators alone. The last column of the table illustrates this even more clearly where addition of just the singles operator to the regular Hamiltonian (S' + $\hat{H}$) lowers the max deviation from 17.6 eV (regular Hamiltonian) to 0.62 eV. However, due to the lack of transcorrelation procedure, the deviations in the ground state energy for this Hamiltonian still remains quite high at around 16 mEh compared to the 21 mEh obtained using the regular Hamiltonian. Thus, the quality of the ground state wavefunction remains to be poor for the S' + $\hat{H}$ Hamiltonian even though the differences in the ground and excited state energies are quite accurate due to error cancellations.
% Results for LiH
\begin{table*}
\captionsetup{justification=raggedright}
\caption{Deviations obtained in the ground state energy (mEh) and the excitation energies (eV) of the six lowest-lying excited states of the LiH molecule using the ANO-RCC-MIN basis set from the reference cc-pVTZ values (second column) for the six different Hamiltonians (see text for details). The superscript in the first column indicates the degeneracy of the given excited state. Parameters used: ($\gamma$, shift) = (0.7, -0.4)}
\begin{ruledtabular}
\begin{tabular}{lccccccc}
{} & {} & {} & {} & {} & {} & {} & {} \\ % blank line
\multicolumn{1}{c}{States} &  \multicolumn{1}{c}{Reference} & \multicolumn{1}{c}{$\hat{H}$} & \multicolumn{1}{c}{$\hat{H}_\text{F12}^{(ij)}$} & \multicolumn{1}{c}{$\hat{H}_\text{F12}^{(pq)}$} & \multicolumn{1}{c}{S' + $\hat{H}_\text{F12}^{(ij)}$} & \multicolumn{1}{c}{S' + $\hat{H}_\text{F12}^{(pq)}$} & \multicolumn{1}{c}{S' + $\hat{H}$}\vspace{0.25cm}\\
\hline
{} & {} & {} & {} & {} & {} & {} & {}\\ % blank line
{} &  \multicolumn{1}{c}{cc-pVTZ}  &  &  &  \multicolumn{2}{c}{ANO-RCC-MB} &  &  \\
\cmidrule(l){3-8}
{} & {} & {} & {} & {} & {} & {} & {}\\ % blank line
S$_0$ & -8.02230 & 29.3 & 15.1 & 15.0 & 8.9  & 8.4 &  27.1  \vspace{0.3 cm} \\
\hline
{} & {} & {} & {} & {}  & {} & {} & {}\\ % blank line
T$_1$       & 3.26 & -0.52  & -0.14  & -0.12  & -0.18   & -0.16  & -0.64 \vspace{0.2 cm} \\
S$_1$       & 3.62 & -0.46  & -0.08  &  -0.10 & -0.09   & -0.11  & -0.54 \vspace{0.2 cm} \\
T$_{2}$(2) & 4.24 & -0.43  & -0.04  &  -0.08 &  0.00   & -0.04  & -0.48 \vspace{0.2 cm} \\
S$_{2}$(2) & 4.61 & -0.47 & -0.07   &  -0.14 &  -0.09  & -0.15  & -0.57\vspace{0.2 cm} \\
T$_3$       & 5.77 &  2.02 &  2.41   &   2.33 &   0.24  & 0.14  &  0.20\vspace{0.2 cm} \\
S$_3$       & 6.37 &  7.06 &  7.46   &   6.44 &   1.65  & 0.55  &  1.17\vspace{0.2 cm} \\
\end{tabular}
\end{ruledtabular}
\label{Tab:LiH}
\end{table*}

\subsection{LiH}
Table~\ref{Tab:LiH} tabulates the deviation in ground state and excitation energies of six lowest-lying excited states of the LiH molecule in an identical layout as the H$_2$ molecule using the same values of $\gamma$ and shift parameters. Here, we have used a minimal basis set ANO-RCC-MB and froze the core electrons in both ground and excited state simulations in order to lower the number of required qubits. The parenthesis in the first column indicates the degeneracy of the excited state. For example, the second singlet and triplet excited states (S$_2$, T$_2$) are doubly degenerate. For the ground state energy, just like before, the transcorrelated Hamiltonians with the singles operator yield the lowest deviations of 8.9 mEh (S' + $\hat{H}_\text{F12}^{(ij)}$) and 8.4 mEh (S' + $\hat{H}_\text{F12}^{(pq)}$) respectively from the reference value. Adding all the orbital pairs to the geminal operator and addition of the singles operator improves the ground state energy the most. However, unlike the H$_2$ molecule, the cc-pVDZ-F12-OptRI basis set was used as the CABS basis in the generation of the transcorrelated Hamiltonian due to the non-availability of the aug-cc-pVTZ-OptRI basis set for Li. We observed that the performance of the cc-pVDZ-F12-OptRI basis is not as optimal compared to the aug-cc-pVTZ-OptRI basis set when ANO-RCC-MB basis set is used as the OBS. Thus, these deviations can be reduced even further with the help of an optimized CABS basis for minimal basis sets. In the case of the excitation energies, the four lowest-lying excited states of the LiH molecule have a weaker dependence on the size of the basis set compared to the $H_2$ molecule, with a max deviation of around 0.5 eV for the regular Hamiltonian. Contrary to the $H_2$ molecule, the regular transcorrelated Hamiltonians (without singles operators) are able to reduce the max deviation to 0.14 eV. The energies of these excited states are thus dominated by the dynamical electron correlation effects. Thus, no major improvements are observed by adding the singles operators for these four states. Also, the performance of both S' + $\hat{H}_\text{F12}^{(ij)}$ and S' + $\hat{H}_\text{F12}^{(pq)}$ Hamiltonians are very similar for these states with a maximum difference of 0.06 eV between the two. However, the deviation for the fifth (T$_3$) and sixth (S$_3$) excited states rises sharply to 2.02 eV and 7.06 eV respectively for the regular Hamiltonian which gets further increased to 2.41 eV and 7.46 eV for the $\hat{H}_\text{F12}^{(ij)}$ Hamiltonian. The $\hat{H}_\text{F12}^{(pq)}$ Hamiltonian decreases these deviations only slightly to 2.33 eV and 6.44 eV respectively. The orbital relaxation effects seem to be very important for these excited states, as can be seen from the last column of table~\ref{Tab:LiH}. Indeed, the S' + $\hat{H}_\text{F12}^{(ij)}$ and S' + $\hat{H}_\text{F12}^{(pq)}$ Hamiltonians are able to bring the max deviation down to 1.65 eV and 0.55 eV respectively. Thus, just like the H$_2$ molecule, the results obtained using the S' + $\hat{H}_\text{F12}^{(pq)}$ are the most accurate for both ground and excited states. 
%(\yz{Since ${\texttt{S'+(F12-H)}_{\texttt{pq}}}$ considers all pairs, I expect it could give better results tan the ${\texttt{S'+(F12-H)}_{\texttt{ij}}}$. But this is not the case for LiH and H2, why?})

% Results for H2O
\begin{table*}
\captionsetup{justification=raggedright}
\caption{Deviations obtained in the ground state energy (mEh) and the excitation energies (eV) of the seven lowest-lying excited states of the H$_2$O molecule using the ANO-RCC-MIN basis set from the reference cc-pVTZ values (second column) for the six different Hamiltonians (see text for details). Parameters used: ($\gamma$, shift) = (1.4, -0.15)}
\begin{ruledtabular}
\begin{tabular}{lccccccc}
{} & {} & {} & {} & {} & {} & {} & {} \\ % blank line
\multicolumn{1}{c}{States} &  \multicolumn{1}{c}{Reference} & \multicolumn{1}{c}{$\hat{H}$} & \multicolumn{1}{c}{$\hat{H}_\text{F12}^{(ij)}$} & \multicolumn{1}{c}{$\hat{H}_\text{F12}^{(pq)}$} & \multicolumn{1}{c}{S' + $\hat{H}_\text{F12}^{(ij)}$} & \multicolumn{1}{c}{S' + $\hat{H}_\text{F12}^{(pq)}$} & \multicolumn{1}{c}{S' + $\hat{H}$}\vspace{0.25cm}\\
\hline
{} & {} & {} & {} & {} & {} & {} & {}\\ % blank line
{} &  \multicolumn{1}{c}{cc-pVTZ}  &  &  &  \multicolumn{2}{c}{ANO-RCC-MB} &  &  \\
\cmidrule(l){3-8}
{} & {} & {} & {} & {} & {} & {} & {}\\ % blank line
S$_0$ & -76.32455 & 342.0 & 151.0 & 121.9 & 33.1  & -1.7 &  233.5  \vspace{0.3 cm} \\
\hline
{} & {} & {} & {} & {}  & {} & {} & {}\\ % blank line
T$_1$ & 7.56 & 3.68 & 5.16 & 3.88 & 1.50 & -0.14 & -0.05 \vspace{0.2 cm} \\
S$_1$ & 8.12 & 4.56 & 6.07 & 4.65 & 1.94 &  0.18 &  0.37  \vspace{0.2 cm} \\
T$_2$ & 9.82 & 3.46 & 4.82 & 3.48 & 1.57 & -0.06 &  0.12 \vspace{0.2 cm} \\
T$_3$ & 9.87 & 4.84 & 6.38 & 4.71 & 1.98 &  0.01 &  0.39\vspace{0.2 cm} \\
S$_2$ &10.14 & 5.16 & 6.53 & 4.99 & 2.33 &  0.25 &  0.66\vspace{0.2 cm} \\
S$_3$ &10.61 & 4.99 & 6.55 & 4.79 & 2.35 &  0.45 &  0.91\vspace{0.2 cm} \\
T$_4$ &11.92 & 3.89 & 5.29 & 3.78 & 1.56 & -0.17 &  0.15\vspace{0.2 cm} \\
\end{tabular}
\end{ruledtabular}
\label{Tab:H2O}
\end{table*}

\subsection{H$_2$O}
The calculations on the H$_2$O molecule also employed ANO-RCC-MB basis set and frozen core settings with the values of $\gamma$ and shift parameter set to 1.4 and -0.15 respectively. From table ~\ref{Tab:H2O}, one can see the regular Hamiltonian is 342 mEh away from the reference value. The $\hat{H}_\text{F12}^{(ij)}$ and $\hat{H}_\text{F12}^{(pq)}$ Hamiltonians are able to reduce it to 151 mEh and $\sim$ 122 mEh respectively. Thus, adding all the orbital pairs in the geminal operator improves the ground state energy by around 29 mEh which is quite significant. After the addition of the singles operator, we obtained deviations of 33.1 mEh and -1.7 mEh for the S' + $\hat{H}_\text{F12}^{(ij)}$ and S' + $\hat{H}_\text{F12}^{(pq)}$ Hamiltonians respectively. Thus, just like in the case of the H$_2$ molecule, we were able to obtain better quality ground state energies than the reference cc-pVTZ values. The excitation energies of all the seven lowest-lying excited states of the $H_2$O molecule, unlike the LiH molecule, have a stronger dependence on the size of the basis set with a max deviation of $\sim$ 5.2 eV for the fifth excited state (S$_2$) using the regular Hamiltonian. The deviations increase even further with the $\hat{H}_\text{F12}^{(ij)}$ Hamiltonian with a max deviation of $\sim$ 6.6 eV, again illustrating the unbalanced treatment of the ground and excited states by this Hamiltonian. Even the $\hat{H}_\text{F12}^{(pq)}$ Hamiltonian doesn't offer any improvement over the regular Hamiltonian and makes the deviations slightly worse for the three lowest-lying excited states (T$_1$, S$_1$, T$_2$) while only minor improvements were noted for the next four (T$_3$, S$_2$, S$_3$, T$_4$). The low-lying excited states of the water molecule is characterized by a mixture of Rydberg and valence correlation effects. In essence, the $\hat{H}_\text{F12}^{(ij)}$ Hamiltonian has the same effect as adding polarization functions for a more accurate description of the valence correlation effects. However, one would require diffuse functions as well in order to accurately describe the Rydberg character of these states. The effect of the singles operator can be seen as an injection of diffusivity into the Hamiltonian. Indeed, the S' + $\hat{H}_\text{F12}^{(pq)}$ Hamiltonian is able to bring down the max deviation from $\sim$ 5.2 eV (regular H) to $\sim$ 0.4 eV. The S' + $\hat{H}_\text{F12}^{(ij)}$ Hamiltonian on the other hand still produces a max deviation of $\sim$ 2.3 eV. Thus, addition of all orbital pairs in the geminal operator is very important for an accurate description of both ground and excited states.
% Results for NH3
\begin{table*}
\captionsetup{justification=raggedright}
\caption{Deviations obtained in the ground state energy (mEh) and the excitation energies (eV) of the eight lowest-lying excited states of the NH$_3$ molecule using the ANO-RCC-MIN basis set from the reference cc-pVTZ values (second column) for the six different Hamiltonians (see text for details). Parameters used: ($\gamma$, shift) = (1.1, -0.30)}
\begin{ruledtabular}
\begin{tabular}{lccccccc}
{} & {} & {} & {} & {} & {} & {} & {} \\ % blank line
\multicolumn{1}{c}{States} &  \multicolumn{1}{c}{Reference} & \multicolumn{1}{c}{$\hat{H}$} & \multicolumn{1}{c}{$\hat{H}_\text{F12}^{(ij)}$} & \multicolumn{1}{c}{$\hat{H}_\text{F12}^{(pq)}$} & \multicolumn{1}{c}{S' + $\hat{H}_\text{F12}^{(ij)}$} & \multicolumn{1}{c}{S' + $\hat{H}_\text{F12}^{(pq)}$} & \multicolumn{1}{c}{S' + $\hat{H}$}\vspace{0.25cm}\\
\hline
{} & {} & {} & {} & {} & {} & {} & {}\\ % blank line
{} &  \multicolumn{1}{c}{cc-pVTZ}  &  &  &  \multicolumn{2}{c}{ANO-RCC-MB} &  & \\
\cmidrule(l){3-8}
{} & {} & {} & {} & {} & {} & {} & {}\\ % blank line
S$_0$ & -56.45043 & 270.6 & 105.3 & 72.6  & 24.3 & -13.4 & 197.7  \vspace{0.3 cm} \\
\hline
{} & {} & {} & {} & {}  & {} & {} & {}\\ % blank line
T$_1$ & 6.06  &  3.27  & 4.66   & 3.24  & 1.85  & 0.13   & 0.42 \vspace{0.2 cm} \\
S$_1$ & 6.62  &  4.06  & 5.49   & 3.86  & 2.44  & 0.53   & 1.00 \vspace{0.2 cm} \\
T$_2$ & 8.21  &  3.69  & 5.14   & 3.47  & 2.12  & 0.16   & 0.67 \vspace{0.2 cm} \\
T$_3$ & 8.21  &  3.70  & 5.15   & 3.48  & 2.13  & 0.16   & 0.68\vspace{0.2 cm} \\
S$_2$ & 8.78  &  4.70  & 6.20   & 4.20  & 2.80  & 0.55   & 1.31\vspace{0.2 cm} \\
S$_3$ & 8.78  &  4.71  & 6.21   & 4.21  & 2.81  & 0.55   & 1.32\vspace{0.2 cm} \\
T$_4$ & 10.92 &  3.38  & 4.46   & 3.20  & 1.18  &-0.38   & 0.08\vspace{0.2 cm} \\
T$_5$ & 10.93 &  3.38  & 4.46   & 3.19  & 1.17  &-0.38   & 0.08\vspace{0.2 cm} \\
\end{tabular}
\end{ruledtabular}
\label{Tab:NH3}
\end{table*}

\subsection{NH$_3$}
For the NH$_{3}$ molecule, the $\gamma$ and shift parameters were chosen to be 1.1 and -0.3 respectively, along with frozen-core settings and ANO-RCC-MB basis set. The trends for the NH$_{3}$ molecule as seen from the table ~\ref{Tab:NH3} are quite consistent with the previous calculations. The S' + $\hat{H}_\text{F12}^{(pq)}$ Hamiltonian again provides ground state energies between cc-pVTZ and cc-pVQZ quality while the regular Hamiltonian yields a deviation of $\sim$ 271 mEh from the cc-pVTZ value. Looking at the table, one can see that three pairs ((T$_2$,T$_3$), (S$_2$,S$_3$), ((T$_4$,T$_5$))) of excited states are nearly degenerate resulting in the appearance of identical deviations with all the Hamiltonians for these states. For the excitation energies, the max deviation produced by the S' + $\hat{H}_\text{F12}^{(pq)}$ Hamiltonian is around 0.5 eV compared to the 4.7 eV and 2.8 eV for the regular and S' + $\hat{H}_\text{F12}^{(ij)}$ Hamiltonians respectively. 
\begin{table*}
\captionsetup{justification=raggedright}
\caption{Estimation of quantum resources required to simulate the ground state of different molecules and basis sets studied in this work using the UCCSD ansatz assuming Jordan-Wigner mapping and frozen-core settings. All excitation operators are considered along with no circuit transpilation or truncation of the Hamiltonian elements}
\begin{ruledtabular}
\begin{tabular}{lccccc}
{} & {} & {} & {} & {} & {}\\ % blank line
\multicolumn{1}{c}{Molecules} &  \multicolumn{1}{c}{Basis} & \multicolumn{1}{c}{Orbitals} & \multicolumn{1}{c}{Qubits} & \multicolumn{1}{c}{Parameters} & \multicolumn{1}{c}{CNOT Gates} \vspace{0.25cm}\\
\hline
{} & {} & {} & {} & {} & {}\\ % blank line
\vspace{0.1 cm}
H$_{2}$     &  6-31G       &   4   & 8   & 15     & 768      \\
{}          &  cc-pVTZ      &  28   & 56  & 783    & 341280    \\
{} & {} & {} & {} & {} & {}\\ % blank line
\hline
{} & {} & {} & {} & {} & {}\\ % blank line
LiH   & ANO-RCC-MB   &   5   & 10  &  24     & 1616      \\
{}    &   cc-pVTZ      &  43   & 86  & 1848    & 1249080  \\
{} & {} & {} & {} & {} & {}\\ % blank line
\hline
{} & {} & {} & {} & {} & {}\\ % blank line
H$_{2}$O    & ANO-RCC-MB   &   6   & 12  &  92     & 8064  \\
{}          & cc-pVTZ      &  57   & 114  & 61904  & 47224272  \\
{} & {} & {} & {} & {} & {}\\ % blank line
\hline
{} & {} & {} & {} & {} & {}\\ % blank line
NH$_{3}$    & ANO-RCC-MB   &  7    & 14   & 204    &  21072   \\
{}         & cc-pVTZ      &  71   & 142  & 98892  & 93571664 \\
\end{tabular}
\end{ruledtabular}
\label{Tab:gate}
\end{table*}

% last table for gate estimation!
\subsection{Quantum resource reduction}
Table \ref{Tab:gate} shows an estimate of the quantum resources required to simulate the ground state of the molecules studied in this work along with the corresponding basis sets assuming Jordan-Wigner mapping and frozen-core settings. All our quantum calculations utilized the UCCSD ansatz and the number of parameters in \ref{Tab:gate} refer to the total singles and doubles excitation operators for the given molecule and basis set. Here, we use the number of CNOT gates as a measure of the quantum circuit complexity. In order to estimate the number of CNOT gates, we used the second-quantized particle-hole formalism for describing the excitation operators and utilized the circuit designs from Ref.~\cite{barkoutsos2018quantum}. The number of CNOT gates required for the exponentiation of a given singles and doubles excitation operator was calculated and summed to obtain the final numbers. These estimates don't take into account any circuit optimization or transpilation and doesn't use any kind of truncation schemes for the Hamiltonian matrix elements. However, they are very useful to describe in a qualitative sense, the massive increase in quantum resource requirements as the number of qubits increases. For example, in the case of H$_2$O and NH$_3$ molecules, going from the minimal basis set of ANO-RCC-MB to the cc-pVTZ basis set results in an increase in the number of CNOT gates by more than 3 orders of magnitude. A smaller number of CNOT gates corresponds to a shallow circuit with lower gate errors making the transcorrelated formalism more suitable for quantum simulations on NISQ devices. Furthermore, we can also reduce the number of measurements in the qEOM procedure quite significantly. The measurement of each single matrix element in the qEOM generalized eigenvalue problem scales as $\mathcal{O}$($N^4$), where N is the number of qubits\cite{pauline_qeom_2020}. For the the NH$_{3}$ molecule, this would translate into a reduction in the number of measurements for a given matrix element by a factor of $10^4$ $((142/14)^4)$.
\section{Conclusions}\label{sec:summary}
We used the canonical transcorrelated theory to construct compact \textit{ab initio} Hamiltonians that can drastically reduce the quantum resources required for accurate simulations of both ground and excited states of molecular systems. In a work by some of the present authors\cite{Motta_CTF12_quantum}, the transcorrelated  Hamiltonians that were obtained through a similarity transformation of the Hamiltonian with an explicitly correlated two-body unitary operator greatly accelerated the recovery of the ground state correlation energies with respect to the size of the basis set. However, the convergence of the excited state properties like excitation energies with these Hamiltonians show a completely different trend and was found to be even slower than the regular Hamiltonian itself. This is not surprising since excited states can have a very different character than the ground state. For example, the low-lying excited states of the water molecule are characterized by a mixture of Rydberg and valence correlation effects but the traditional explicitly correlated methods can only capture the missing dynamical electron correlation. Also, the previous formalism was not able to recover the missing dynamical correlation effects between electrons in occupied and virtual orbitals due to the absence of orbital pairs involving virtual orbitals in the definition of the two-body geminal operator. In this work, we have addressed all these points by re-defining the two-body geminal operator to include all orbital pairs and added a singles operator in the similarity transformation procedure to account for the orbital relaxation effects, resulting in a balanced treatment of both ground and excited states. The new transcorrelated Hamiltonians can produce ground state energies comparable to the cc-pVTZ basis, even with a minimal basis set. Furthermore, it can reduce the errors in the excitation energies by more than an order of magnitude. This can potentially lead to more than one and three orders of magnitude reduction in the number of qubits and CNOT gates respectively, in the VQE procedure. Consequently, the quantum simulations with the transcorrelated Hamiltonian are expected to be more noise resilient on NISQ devices. 

\begin{acknowledgements}
Research presented in this article was supported by the Laboratory Directed Research and Development (LDRD) program of Los Alamos National Laboratory (LANL) under project number 20200056DR. LANL is operated by Triad National Security, LLC, for the National Nuclear Security Administration of U.S. Department of Energy (contract no. 89233218CNA000001). We thank LANL Institutional Computing (IC) program for access to HPC resources. AK acknowledges the help of Tanvi Gujarati (IBM, USA) for help in quantum resource estimations.
\end{acknowledgements}

%\clearpage
\bibliography{ref}
\end{document}